\documentclass[useAMS,usenatbib,letterpaper]{mn2e}
\usepackage[totalwidth=480pt, totalheight=680pt]{geometry}
\usepackage{graphicx}
\usepackage{geometry,natbib}

\def\deg{^{\circ}}
\def\etal{{\it et al.}\thinspace}

\def\eg{{\it e.g.,}\thinspace}
\def\eq{\begin{equation}}
\def\en{\end{equation}}

\def\etal{{\it et al.}\thinspace}

\def\eg{{\it e.g.}\thinspace}
\def\apj{{\it Ap.J.}\thinspace}

\def\aap{{\it A\&A}\thinspace}
\def\mnras{{\it MNRAS}\thinspace}
\def\P3hat{{\mathaccent 94 P}_3}

\def\nat{{\it Nature}\thinspace}

\title[Deep Analyses of Nulling in Arecibo Pulsars Reveal Further Periodic Behavior]{Deep Analyses of Nulling in Arecibo Pulsars Reveal Further Periodic Behavior}
\author[Jeffrey L. Herfindal \& Joanna M. Rankin]{Jeffrey L. Herfindal and Joanna M. Rankin  \\
Physics Department, University of Vermont, Burlington, VT 05405\thanks{J.Herfindal@gmail.com; Joanna.Rankin@uvm.edu} }

\begin{document}

\date{Accepted 2008 month day. Received 2008 month day; in original form 2008 month day}

\pagerange{\pageref{firstpage}--\pageref{lastpage}} \pubyear{2008}

\maketitle

\label{firstpage}

\begin{abstract}
Sensitive Arecibo observations provide an unprecedented ability to detect 
nulls for an accurate pulse-modulation quelling (PMQ) analysis.  We 
demonstrate that a number of conal pulsars show ``periodic nulling'' similar 
to the phenomenon found earlier in pulsar B1133+16.  
\end{abstract}

\maketitle

\section{Introduction}
The complex inner workings of the pulsar emission mechanism still remain 
something of a mystery four decades after their discovery.   Pulsar emission 
is both difficult and fascinating, in part, because of its prominent modulation 
phenomena---in particular the ``big three'' effects of subpulse drifting, ``mode'' 
changing, and the pulse ``nulling'' that is the subject of this paper.

The nulling phenomenon has been very perplexing since first identified by 
Backer (1970), because the nulls affected all components, even interpulses, 
and were simultaneous at all frequencies.  Nulls thus appeared to represent 
a temporary cessation of the pulsar emission process, but strangely ``memory'' 
was observed across nulls in some cases (Page 1973).  Null fractions (NF) 
were computed for many of the known pulsars, and found to range from less 
than 1\% up to 70\% or so; but some half the stars appear not to null at all.  The 
first systematic study of such null fractions was made by Ritchings (1976), 
who found a correlation between the NF and a pulsar's spindown age $\tau$.  
Ten years later, Rankin (1986) showed that while old pulsars null 
more than young ones, many old pulsars do not null at all.  

Ever more dramatic cases of extreme nulling---that is, apparent episodes of 
activity and inactivity---have been discovered over the intervening two decades: 
\eg, those of B0826--34 last for hours (Durdin \etal\ 1979); B1944+17 nulls 70\% 
of the time (Deich \etal\ 1986); and B1931+24 seems to cycle semi-periodically 
on a time scale of a few weeks (Kramer \etal\ 2006).

Evidence has steadily accrued over the last few years that the nulls 
in many pulsars are not random turn-offs:  the subpulse ``memory'' across nulls 
in B0809+74 closely associates them with the star's drift (van Leeuwen \etal\ 
2002, 2003); evidence of sputtering emission during nulls has been identified 
in both B0818--13 (Janssen \& van Leeuwen 2004) and B1237+25 (Srostlik \& 
Rankin 2005); almost all the nulls occur in one mode in B2303+30 (Redman 
\etal\ 2005); in B0834+06 the nulls tend to occur on the weak phase of the 
star's alternate-pulse modulation cycle (Rankin \& Wright 2007a); and finally 
the nulls in J1819+1305 exhibit a strong 57-stellar-rotation-period cyclicity 
(Rankin \& Wright 2007b).  These circumstances strongly suggest that nulls, 
in many cases, represent ``empty'' sightline traverses through a regularly 
rotating ``carousel'' subbeam system (\eg, Deshpande \& Rankin 2001).  

In a recent paper (Herfindal \& Rankin 2007; hereafter Paper I), we identified 
evidence for periodic nulling in pulsar B1133+16, a pulsar which exhibits no 
regular subpulse modulation.  The nulls could be distinguished with great 
certainty in this star, and we then applied a straightforward method of filling 
the non-null pulses with the appropriately scaled-down average profile.  This 
pulse-modulation quelling (hereafter PMQ) technique then confirmed that  
star's low frequency modulation feature was associated with its nulls.  Here, 
the implication is that this star's nulls are produced by a relatively stable, but 
irregular and sparsely filled carousel-beam system whose rotation gives a 
rough periodicity to ``empty'' sightline passes.  Such an interpretation 
may also explain Bhat \etal's (2007) result that B1133+16's null pulses are 
not strictly simultaneous at all frequencies.  They found about a 5\% excess 
of nulls at meter wavelengths, and this may result from the star's larger conal 
emission pattern here.

These various current results lend new importance to understanding pulsar 
nulling more fully.  Carousel-related ``pseudonulls'' may occur widely and 
explain much, but in certain stars the evidence is very strong that their nulls 
represent a cessation of their emission (\eg, B1931+24), so at least 
two different types of nulls are implied.  Nulling is also closely associated 
with mode changing (\eg, Wang \etal\ 2007), and the recently discovered 
rotating-radio transients (RRATs) naturally raise questions about the nature 
of such pulsars' long dormancies between their sporadic powerful bursts 
(McLaughlin \etal\ 2005).  

Here we continue the analytical effort begun in Paper I---that is, applying 
the PMQ method to a small population of pulsars with conal profiles in order 
to test its efficacy and interpret its results in a larger context.  Our Arecibo 
observations and analysis methods are briefly discussed in \S 2 and our 
results for each pulsar in  \S 3.  In \S 4 we summarize and discuss the 
results overall.  

\section{Observations \& Methods}
All of the observations were carried out using the 305-meter Arecibo 
Telescope in Puerto Rico.  Observations were conducted in the P band 
at 327 MHz.  They used the same correction methods and instrumental 
techniques as in Paper I.  Table~1 gives the resolution, length, and date 
of each observation.

Longitude-resolved fluctuation spectra (hereafter LRF) of the total power 
component (Stokes I) were computed for all of the observations in Table 1 
using Fourier transforms of length 256\footnote{Pulsar B0525+21, 
observation dated 2003 October 4, used a FFT of length 128.  Pulsars 
B2303+30 and J0540+32 both used a FFT of length 512.}.  
Figure~\ref{Fig.2} shows the LRF spectra for pulsar B2034+19 with the 
aggregate intensity (middle panel of the right column) showing a clear 
$57\pm$6 $P_1$ feature.\footnote{$P_1$ is that particular pulsar's period.}

\begin{figure}
\includegraphics[width=78mm]{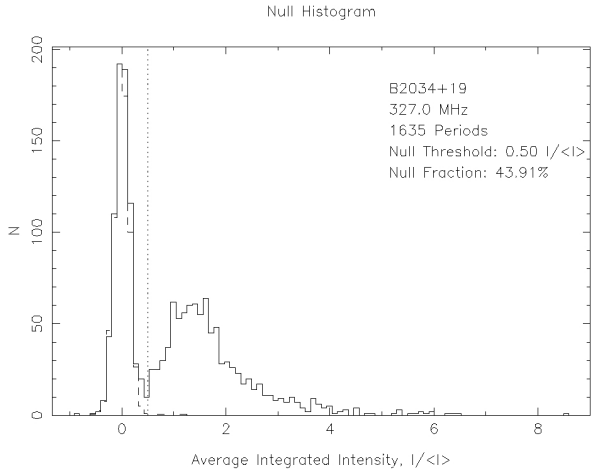}
\caption{Null histogram for the pulsar B2034+19.  The integrated-intensity 
distribution of the pulses (solid line) and that of the off-pulse region (dashed 
line) are plotted.  The vertical dotted line represents an integrated intensity
limit of 0.50 $<$$I$$>$ to distinguish the nulls.  Notice the null distribution is 
contiguous with that of the pulses frustrating any attempts to decisively 
measure the null fraction.  Forty one (41) pulses were ignored due to bad 
baselines.}
\label{Fig.1}
\end{figure}

\begin{figure}
\includegraphics[width=78mm]{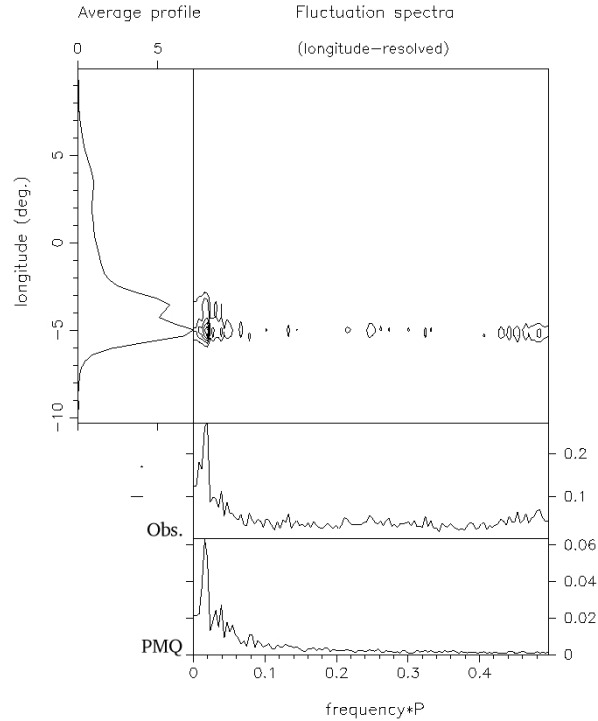}
\caption{Typical LRF plot for pulsar B2034+19 computed in total power 
(Stokes I).  The main panel gives the spectra according to the average 
profile in the left panel, and the integrated spectrum is shown in the 
middle panel, right column.  LRF spectra of the artificial ``PMQ'' PS 
corresponding to this pulsar is in the bottom panel.  Most pulse modulation 
was quelled by substituting a scaled-down average profile for the pulses 
while zero intensity was substituted for detected nulls (see text).  Here, 
as for most of the observations, an FFT length of 256 was used.}
\label{Fig.2}
\end{figure}

The null histogram for pulsar B2034+19 is shown in Figure~\ref{Fig.1}.  Notice
 the strong presence of pulses with zero (or near zero) aggregate intensity.  
 However, note that the distribution is continuous between the pulses and 
 nulls, frustrating any possibility of delineating the two populations positively.  
 The dotted line represents the optimal boundary between nulls and pulses, 
 corresponding to PSR B2034+19; 44\% of the pulses fall below this threshold.  
 Suitable pulse-intensity histograms were plotted for each observation to 
 determine a plausible null threshold.

\begin{table}
\begin{center}
\caption{Observational parameters.}
\begin{tabular}{cccc}
\hline
\hline
Pulsar & Date & Length &Resolution \\
     & (m d$^{-1}$ yr$^{-1}$) & (pulses) &($\deg$/sample) \\
\hline
B0045+33 & 01/07/2005 & 1085 & 0.30 \\
B0301+19 & 01/08/2005 & 1729 & 0.19 \\
B0525+21 & 10/04/2003 & 636   & 0.35 \\
		 & 10/07/2006 & 961$^a$  &  0.35 \\
J0540+32 &  10/07/2006 & 1145 & 0.48 \\
B0751+32 & 10/04/2003 & 1248 & 0.35 \\
	           & 10/07/2006 & 2080 & 0.35 \\
B0823+26 & 10/04/2003 & 3392 & 0.35 \\
B0834+06 & 10/05/2003 & 3789 & 0.35 \\
	           & 05/06/2006 & 1920 & 0.35 \\
B1237+25 & 07/12/2003  & 2340 & 0.35 \\
		  & 07/13/2003 & 5094 & 0.35 \\
		  & 07/20/2003 & 4542 & 0.35 \\
		  & 01/08/2005 & 5209 & 0.13 \\
J1649+2533& 01/06/2005 & 1044 & 0.36 \\
		    & 02/12/2006 & 2818 & 0.36 \\
J1819+1305 & 02/12/2006 & 3394 & 0.51 \\
B1831-00   & 01/07/2005 & 1151 & 0.71 \\
B1839+09  & 01/07/2005 & 1573 & 0.48 \\
B1848+12 & 10/19/2003 & 2074 & 0.15 \\
		  & 08/19/2006 & 1037 & 0.48 \\
B1918+19 & 02/12/2006 & 3946 & 0.39 \\
B2034+19 & 01/07/2005 & 1676 & 0.36 \\
B2122+13 & 01/08/2005 & 1038 & 0.36 \\
B2303+30 & 10/07/2003 & 1526 & 0.23 \\	        
B2315+21 & 10/07/2003 &  622 & 0.35 \\
		 &  01/07/2005 & 2491 & 0.26 \\ 
\hline
\end{tabular}
\end{center}
\scriptsize
$^a$ The last 121 pulses of this observation were ignored due to noticeable 
interference.
\normalsize 
\label{T1}
\end{table}

Pulse modulation quelling (hereafter PMQ) was preformed on each observation 
by computing a binary series of nulls and pulses, particular to each observation.  
To get this, the null threshold from the pulse-intesity histograms was compared 
to each pulse, within the same window, in order to determine if it was a pulse 
or null.  A new pulse sequence was created corresponding to the natural pulse 
sequence by substituting a scaled-down average profile for pulses and zero 
intensity for the ``null'' pulses.  The lowest panel of the right column in Fig.~\ref{Fig.2} 
shows the LRF of the PMQ pulse sequence for pulsar B2034+19.  Notice that the 
same low frequency feature persists!  

\section{Individual Pulsar Parameters}

\textbf{B0301+19:}
This pulsar has been found to have straight drift bands in both components 
on the pulse--stack by Sch\"onhardt \& Sieber (1973).  Weltevrede \etal\ 
(2006, 2007; hereafter W0607), during their two-dimensional Fourier series 
(hereafter 2DFS) analysis, found the trailing component exhibits a broad drift 
feature with a much higher $P_3$ value than in the leading component. The 
LRF shows two prominent low frequency features which have a much higher
value then any of the $P_3$ values reported.  The highest feature is only 
prominent in the trailing component (this may be due to the sporadic nature 
of the trailing component) with the $51\pm5$ $P_1$ feature appearing in both 
components.  After PMQ analysis only the 51 period feature remains.  Rankin 
(1986) found a null fraction around 10\%.  The null histograms show a slightly 
larger null fraction, on order of 13\%.  \\

\textbf{B0525+21:}
Backer (1973) found if emission occurs in component II then it can be 
predicted, reliably, that the next subpulses will be in component I.  He 
also found that the most prominent feature is at 0.025 cycles/$P_1$ 
(40 $P_1$) and that it is present at all longitudes across the pulse.  
Taylor \etal\ (1975) found a very weak preference for negative subpulse 
drift in adjacent pulses, but it is almost equal to a positive subpulse drift.  
Recently W0607 found the trailing component shows a drift feature with 
opposite drift direction (at 21-cm), while there is evidence for a preferred 
negative drift direction in the leading component at 21-cm and 92-cm.  
They also confirmed a $P_3$ around 3.8 $P_1$.  The LRF features vary 
between the observations with only one remaining feature after PMQ 
analysis.  The 4-odd $P_1$ drift feature is only seen in the longer of the 
observations in the normal LRF.  The null histograms show a continuous 
distribution between that of the pulses and the nulls, giving a varying null 
fraction between the observations.  These null fractions are consistent 
with Ritchings (1976) of 25\%.

\textbf{J0540+32:}
This newly discover pulsar was found during the Arecibo Pulsar survey 
using the ALFA (Cordes \etal\ 2006).  This pulsar has a clear separation 
between the null and pulse distributions with 54\% percent of the pulses 
falling below our null threshold.  There is a distinct very low frequency 
feature at $256 \pm$64 $P_1$.  The PMQ spectra shows the same feature 
with another harmonic feature at $85 \pm$7 $P_1$.  

\textbf{B0751+32:}
At 430 MHz Backus (1981) identified a preference for negative drift in this 
pulsar; recently, W0607 found the leading and trailing components show 
a low frequency feature at $60\pm$20 $P_1$ and $70\pm$10 $P_1$, respectfully 
(at 92-cm).  Once the observations underwent Fourier transforms, the LRF's 
revealed that the low frequency feature changes slightly between our 
observations.  Nevertheless, after PMQ analysis the main features in the 
original LRF's remain.  Nulling in B0751+32 occurs around 34\% 
(Backus, 1981).  The null fraction from the pulse-intensity distribution 
histograms changes slightly but is consistent with Backus's results.  

\textbf{B0834+06:}
Taylor \etal\ (1969) found a strong response at 0.462 cycles/$P_1$ (2.16 $P_1$), 
confirmed by Slee \& Mulhall (1970), Sutton \etal\  (1970) identified it as a 
drifting feature, Backer (1973), Taylor (1975) found a preference for positive 
subpulse drift, and recently by Asgekar \& Deshpande (2005) and W0607 
confirmed the drifting feature's measurement of $2.2 P_1$.  Rankin \& Wright 
(2007a) showed the nulls are not randomly distributed, occurring in a periodic 
matter.  The LRFs of these observations have a clear feature around 2.16 
$P_1$ which is conserved after PMQ.  An unexpected feature was found after 
PMQ analysis of the longer observation at 16 $P_1$.  A null fraction of around 
7.1\% was found by Ritchings (1976); Rankin \& Wright (2007a) found that no 
more then 9\% of pulses are nulls.  These observations have very distinct 
null--pulse distributions resulting in accurate null fractions which confirm these 
previous results.


\textbf{J1649+2533:}
This pulsar was found to have a null fraction of 30\% by Lewandowski \etal\
(2004).  The null fractions in these observations are slightly lower (on order
of 21\%).  Lewandowski \etal\ also measured a $P_3$ = 2.2 $P_1$; both of the
observations show this feature towards the outer edges of the profile, with a
clear separation in the middle.  After PMQ analysis, only the 27-odd $P_1$
feature remains.

\begin{table*}
\begin{center}
\caption{Observed low frequency feature(s), features after PMQ, and null fractions.}
\begin{tabular}{llcccccc}
\hline
J2000 & B1950 & Length    & NF             & LRF Feature(s)   & PMQ Feature & Figure \\
 name & name	 & (pulses)  &  (per cent) &     ($P_1$)           &	($P_1$)	  &\\
\hline
\hline
 J0304+1932 & B0301+19& 1729& 15      &$128 \pm 32$ & ---                 &Fig.~A.\ref{Fig0301} \\
		      &                    &          &             &$ 51\pm5$      & $51\pm5$    & \\
 J0528+2200 & B0525+21 & 636 & 22     & $51 \pm 10$  & $43 \pm 7$   & Fig.~A.\ref{Fig0525}\\
 		      &	                     & 961 & 32    & $39 \pm 3$    &  ---                  & \\
		      &	                     &	    &           & $ 28 \pm 2$    &    ---	       & \\
		      &	                     &	    &           & $ 22 \pm 1$    & $ 23 \pm 1$ & \\	   
		      &	                     &	    &           & $ 4.57 \pm 0.04$ & ---              & \\	
J0540+32	   & 		---	& 1145 & 54    & $256 \pm 64$ & $256 \pm 64$ &Fig.~A.\ref{Fig0540} \\
J0754+3231 &B0751+32 & 1248 & 36   &$ 51 \pm 5$ & $ 51 \pm 5$   & Fig.~A.\ref{Fig0751}   \\ 
	              &		         & 2080 & 40   &$ 64 \pm 8$ & $ 73 \pm 10$ &   \\
&&&&&&\\
J0837+0610 & B0834+06 & 3789 & 9    & $2.17 \pm 0.01$ & $2.18 \pm 0.01$& Fig.~A.\ref{Fig0834} \\
		     &		          &            &          & ---                          & $16 \pm 1$ & \\ 
   		      &                    & 1920 & 9    & $2.15 \pm 0.01$ & $2.15 \pm 0.01$&  \\         
J1649+2533 & --- & 1044 & 21    & $64 \pm 8$ & --- 			&Fig.~A.\ref{Fig1649} \\
		     &&	          &         & $28 \pm 2$ & $28 \pm 2$    & \\
		     &&	          &         & $2.5 \pm 0.1$ & ---           & \\  
		     &      & 2818 & 27   & $57 \pm 6 $ & $57 \pm 6 $ &\\
		     &	    &		&  	    & $26 \pm  1$ & $26 \pm 1$	    & \\
		     &&           &                & $2.5 \pm 0.1 $ & ---		  & \\
J1819+1305 & --- & 3394 & 46 &$57\pm6$ & $64\pm 8 $ &Fig.~A.\ref{Fig0540} \\
J1841+0912 & B1839+09 & 1573 & 3  & $37\pm 3$ & $37\pm 3$ & Fig.~A.\ref{Fig0301} \\
		     &		          &	      &            & $28\pm 2$ & ---                 & \\     
&&&&&&\\
J1921+1948 & B1918+19 & 3946 & 9     &$85\pm 14$ & $85\pm 14$ & Fig.~A.\ref{Fig1918} \\
                         &                    &	      &          & ---	                & $43\pm 4$   & \\
J2037+1942 & B2034+19 & 1676 & 44   & $57\pm6$ & $57\pm6$ & Figs.~\ref{Fig.1},\ref{Fig.2}\\
J2305+3100 & B2303+30 & 1526 & 11  & $102 \pm 20$ &  $128 \pm 32$  &Fig.~A.\ref{Fig1918} \\
			&		 &	    &  		   & ---                     & $ 64 \pm 8$       & \\
			&		 &	    &  		   &   $37 \pm 3$  & $37 \pm 3$         & \\			
\hline
\end{tabular}
\end{center}
\label{table2}
\end{table*}

\textbf{J1819+1305:}
A very clear feature at $57\pm$6 $P_1$ in the LRF is spread across the whole 
pulse with a slight absents in the middle.  This feature remains as prominent 
after PMQ analysis.  The null histogram is a continuous distribution deterring 
any possibility of an accurate measurement of ``null'' pulses.


\textbf{B1839+09:}
W0607 confirmed the result (at 21 and 92 cm) found by Backus (1981); stating
that there is no preference in the drift direction for the subpulse modulation.
The LRF for this observation shows a clear feature at 37$\pm$3 $P_1$,
which also remains after PMQ was performed.  The null histogram shows
clearly that the pulse distribution trails off into the null distribution, resulting in a
null fraction of 2\%.

\textbf{B1918+19:}
Four drifting modes identified by Hankins \& Wolszczan (1987).   This pulsar has 
a continuous distribution between the ``null'' and pulse distributions in the pulse-
energy histogram.  A conservative null threshold was chosen, correspondingly 
9\% of these pulses fall below this threshold.  The LRF shows a strong low-
frequency feature at 85$\pm$14 $P_1$.  This feature persists after PMQ analysis.

\textbf{B2034+19:}
This pulsar has a clear null and pulse distribution in the pulse-intensity histogram.  
The LRF shows a clear feature at $57\pm$6 $P_1$ which remains after PMQ 
analysis.

\textbf{B2303+30:} 
Redman \etal\ (2005) found two modes in this pulsar with almost all of the nulls come
about in the `Q' (quiescent) mode.  11\% of the pulses in this pulsar fall below the null
threshold in the pulse-energy histograms.  The LRF shows two low frequency features:
one at 102$\pm$20 $P_1$ with the other at 37$\pm$3 $P_1$.  Both of these features 
remain after PMQ analysis.

\begin{table}
\begin{center}
\caption{Observed null fractions of ``featureless'' pulsars.}
\begin{tabular}{llcccccc}
\hline
J2000 & B1950 & Length    & NF      \\
 name & name	 & (pulses)  &  (per cent)\\
\hline
\hline
 J0048+3412 & B0045+33& 1085 & 22  \\
J0826+2637 & B0823+26 & 3392 & 7 \\
J1239+2453 & B1237+25 & 2340 & 5 \\
		     &		          & 5094 & 6 \\
		     &		          & 4542 & 6 \\	
		     &	                   & 5209 & 5 \\
J1834-0010& B1831--00 & 1151 & 3 \\
J1851+1259 & B1848+12 & 2074 & 54 \\
		     &			& 1037 & 50 \\
J2124+1407 & B2122+13 & 1038 & 27 \\
J2317+2149 & B2315+21 & 622 & 2 \\
                        &		         & 2491 & 2 \\
	     
\hline
\end{tabular}
\end{center}
\label{table2}
\end{table}

\section{Discussion}

Our surprise in Paper I was that the PMQ analysis associated B1133+16's low frequency 
feature so clearly with its {\em nulls}!  In this larger effort, we are no longer surprised to 
find abundant evidence for null-related periodicities in this population of conal dominated 
pulsars.  We do, however, find pulsars whose nulls show no obvious periodicity as well 
as cases where PMQ reveals several, probably harmonically related, periodicities.  

Specifically, the PMQ analysis did not always identify null periodicity.  Table 3 lists seven 
pulsars which clearly have a null fraction but do not, except for B1237+25 \& B1831--00, 
have a clear LRF feature.  These seven pulsars tend to produce either very broad 
features or featureless (``random'') PMQ LRF spectra.  These non-definitive results could 
come about because a pulsar:  i) has several different drift modes; ii) an irregular carousel 
rotation rate; iii) an unfavorable sight-line traverse across the edge of the carousel; or 
iv) has nulls that are completely random.  

In all cases the pulse-energy distributions were continuous with those of the null distributions.  
Strangely, we have yet to encounter any example of a pulsar whose pulses and nulls are fully 
disjoint.

The results of this paper amplify the evidence reviewed above to the effect that many pulsar 
nulls are neither random nor complete turn-offs.  Other recent evidence (\eg, B1931+24), 
however, all but confirms absolutely that some pulsar nulls do represent a complete or 
almost complete cessation of the emission.  The conclusion then can hardly be escaped 
that pulsar nulls are of at least two distinct types.  And the recently discovered RRATs then 
reverse the traditional null question:  ``How can an electrodynamic system which is almost 
always in the null state then flash very occasionally into brilliance?''

\section*{Acknowledgments}
Portions of this work were carried out with support from US National Science Foundation Grant AST 99-87654.  Arecibo Observatory is operated by Cornell University under contract to the US NSF.  This work used the NASA ADS system.  

{}
\bsp
\label{lastpage}


\begin{thebibliography}{99}
\bibliographystyle{plainnat}
\bibitem[Asgekar \& Deshpande] ]{AD05} Asgekar, A., \& Deshpande, A.A, \mnras, 2005, 357, 1105
\bibitem[Backer 1970]{B70} Backer, D.C., 1970, \nat, 228, 42
\bibitem[Backer 1973]{BA73} Backer, D. C. 1973, \apj, 182, 245
\bibitem[Backus, P. R. 1981]{B81} Backus, P. R. 1981 Ph.D. Thesis, The University of Massachusetts
\bibitem[Bhat \etal\ 2007]{B07} Bhat, N.D.R., Gupta, Y., Kramer, M., Karastergiou, A., Lyne, A.G., 
	\& Johnston, S. 2007, \aap, 462, 257
\bibitem[Cordes \etal\ 2006]{C06} Cordes, J.M., Freire, P.C.C., Lorimer, D.R., \etal\, 2006, \apj, 637, 446
\bibitem[Deshpande \& Rankin 2001]{DR01} Deshpande, A.A., Rankin, J.M., 2001, \mnras, 322, 438 
\bibitem[Deich \etal\ 1986]{} Deich, W.T.S., Cordes, J. M., Hankins, T. H., \& Rankin, J. M. 1986, 
	\apj, 300, 540
\bibitem[Durdin \etal\ 1979]{} Durdin, J. M., Large, M. I., Little, A. G., Manchester, R. N., Lyne, A. G., 
	\& Taylor, J. H. 1979, \mnras, 186, 39
\bibitem[Hankins \& Wolszczan 1987]{HW87} Hankins, T.H. \& Wolszczan, A., 1987, \apj, 318, 410
\bibitem[Herfindal \& Rankin 2007]{H07} Herfindal, J. L., Rankin, J. M. 2007 \mnras, 380, 430
\bibitem[Janssen \& van Leeuwen 2004]{J04} Janssen, G.H., van Leeuwen, A.G.J., 2004, \aap, 425, 255
\bibitem[Kramer \etal\ (2006)]{K06} Kramer, M., Lyne, A. G.,  OНBrien, J. T.,  Jordan, C. A., \& 
	Lorimer, D. R. 2006, Science, 312, 549
\bibitem[Lewandowski \etal\ 2004]{L04} Lewandowski, W., Wolszczan, A., Feiler, G., \etal\ 2004, \apj, 600 905
\bibitem[van Leeuwen et al 2002]{L02} van Leeuwen, A.G.J., Kouwenhoven, M.L.A., 
	Ramachandran, R., Rankin, J. M., \& Stappers, B. W.  2002, \aap, 387, 169
\bibitem[van Leeuwen et al 2002]{L02} van Leeuwen, A.G.J., Stappers, B. W., Ramachandran, R., 
	\& Rankin, J. M.  2003, \aap, 399, 223
\bibitem[mllk]{MLLK06} McLaughlin, M.A., Lyne, A.G., Lorimer, D.R., Kramer, M., \etal\, 2006
	\nat, 439, 817
\bibitem[Page 1973]{P73} Page, C.G., 1973, \mnras, 163, 29
\bibitem[Rankin 1986]{R86} Rankin, J.M., 1986, \apj, 301, 901
\bibitem[Rankin \& Wright (2007a)]{rw07} Rankin, J. M. \& Wright, G.A.E. 2007a, \mnras, 379, 507
\bibitem[Rankin \& Wright (2007b)]{rw08} Rankin, J. M. \& Wright, G.A.E. 2007b, \mnras, in press
\bibitem[Redman \etal\ (2005)]{RWR05} Redman, S.R., Wright, G.A.E., Rankin, J.M., 2005, 
	\mnras, 357, 859
\bibitem[Ritchings 1976]{R76} Ritchings, R.T. 1976, \mnras, 176, 249
\bibitem[Sch\ddot{o}nhardt \& Sieber 1973]{S73} Sch\"onhardt, R. E. \& Sieber, W. 1973 Astrophys. Lett., 14, 61
\bibitem[Slee \& Mulhall 1970]{S70} Slee, O.B., Mulhall, P.S., 1970,  Proc. Astr. Soc. Australia, 1, 322
\bibitem[Srostlik 2005]{SR05} Srostlik, Z. \& Rankin, J.M., 2005, \mnras, 362, 1121
\bibitem[Sutton \etal\ 1970]{Su70} Sutton J. M., Staelin, D. H., Price, R. M., \& Weimer, R. 1970, \apj, 159, 89
\bibitem[Taylor \etal\ 1969]{T69} Taylor, J. H., Jura, M., Huguenin, G. R. 1969 \nat, 223, 797
\bibitem[Taylor \etal\ 1975]{T75} Taylor, J. H., Manchester, R. N., Huguenin, G. R. 1975, \apj, 195, 513
\bibitem[Wang \etal\  2007]{wmj07} Wang, N., Manchester, R. N., \& Johnston, S. 2007, 
	\mnras, 377, 1383
\bibitem[Weltevrede \etal\ 2006]{WES06} Weltevrede, P.,  Edwards, R. T., \& Stappers, B.  2006, 
	\aap, 445, 243
\bibitem[Weltevrede \etal\ 2007]{WSE07} Weltevrede, P., Stappers, B. W., \& Edwards, R. T. 2007, 
	\aap, 469, 607
\end{thebibliography}
\end{document}